\documentclass[12pt, preprint, flushrt, mathptmx]{aastex}

\usepackage{epsf}
\usepackage{verbatim}

\newcommand{\ha}{H$\alpha$}
\newcommand{\comments}[1]{}

\begin{document}

\sffamily

\title{Dynamics in Sunspot Umbra as Seen in New Solar Telescope and Interface Region Imaging Spectrograph Data}
\author{V. Yurchyshyn$^{1,2}$, V. Abramenko$^{1,3}$, and A. Kilcik$^{4}$ }

\affil{\it $^1$Big Bear Solar Observatory, New Jersey Institute of Technology, \\ Big Bear City, CA 92314, USA}
\affil{\it $^2$Korea Astronomy and Space Science Institute, \\ 776 Daedeok-daero, Yuseong-gu, Daejeon, 305-348, Korea}
\affil{\it $^3$Pulkovo Observatory, Russian Academy of Sciences,  \\ Pulkovskoe sh. 65, St. Petersburg, 196140, Russia}
\affil{\it $^4$Department of Space Science and Technologies, Akdeniz University, \\ 07058 Antalya, Turkey}

\begin{abstract}
We analyse sunspot oscillations using Interface Region Imaging Spectrograph (IRIS) slit-jaw and spectral data and narrow-band chromospheric images from the New Solar Telescope (NST) for the main sunspot in NOAA AR 11836. We report that the difference between the shock arrival times as measured the Mg II k 2796.35\AA\ and Si IV 1393.76\AA\ line formation levels changes during the observed period and peak-to-peak delays may range from 40~s to zero. The intensity of chromospheric shocks also displays a long term (about 20~min) variations. NST's high spatial resolution \ha\ data allowed us to conclude that in this sunspot umbral flashes (UFs) appeared in the form of narrow bright lanes stretched along the light bridges and around clusters of umbral bright points. Time series also suggested that UFs preferred to appear on the sunspot-center side of light bridges, which may indicate the existence of a compact sub-photospheric driver of sunspot oscillations. The sunspot's umbra as seen in the IRIS chromospheric and transition region data appears bright above the locations of light bridges and the areas where the dark umbra is dotted with clusters of umbral dots. Co-spatial and co-temporal data from the Atmospheric Imaging Assembly on board Solar Dynamics Observatory showed that the same locations were associated with bright footpoints of coronal loops suggesting that the light bridges may play an important role in heating the coronal sunspot loops. Finally, the power spectra analysis showed that the intensity of chromospheric and transition region oscillations significantly vary across the umbra and with height, suggesting that umbral non-uniformities and the structure of sunspot magnetic fields may play a role in wave propagation and heating of umbral loops. 
\end{abstract}

\comments{Key words: Sun: atmosphere -- Sun: chromosphere -- Sun: oscillations}

\section{Introduction}

The sunspot oscillations, as measured in the photosphere, show broad range of frequencies that peak at  3~ mHz (period of approx. 5~min).  They are generally coherent over a large fraction of the sunspot umbra and five minute p-modes may propagate completely across the entire sunspot \citep[e.g.,][]{2004A&A...424..671K}. Moreover, there are indications that the oscillations proceed in 20 -- 40~min long wave trains \citep{2008ASPC..383..279B,2013A&A...554A.146K} the origin of which is not well understood. Oscillations of the sunspot umbra as measured chromospheric lines show major peak around 5 mHz (period of 3~min) and amplitudes of several kilometers per second.  They are thought to represent various magneto-hydrodynamic waves propagating within the sunspot and thought to be generated by sunspot oscillations \citep[e.g.,][and references therein]{2000SoPh..192..373B,2013SoPh..288...73M}. There is a small time lag in the oscillatory patterns measured at different altitudes, which is consistent with the general picture of waves propagating up from the photosphere to the base of the corona \citep[e.g.,][]{Tian_2014}. Although the energy flux associated with with 3~min oscillations appears be too low to maintain the observed temperature of coronal loops \citep{2005RSPTA.363.2743D}, \cite{2000A&A...362.1151N} and later \cite{2001A&A...379.1106T} showed that when the entire spectrum of propagating slow magneto-acoustic waves is considered then they could provide energy flux sufficient to account for heating of active region loops. For comprehensive review of sunspot oscillations reader is referred to papers by \cite{2006RSPTA.364..313B} and \cite{0004-637X-719-1-357}.

It was found that the intensity of coronal loops associated with plage areas display periods of the order of 321 $\pm$ 74 s, whereas sunspot loops oscillate with periods of the order of 172 $\pm$ 32 s, which indicates that photospheric oscillations above a sunspot have the ability to propagate through the transition region (TR) and into the corona \citep{2002A&A...387L..13D}. \cite{2013ApJ...762...42S} suggested that the propagating waves may modulate magnetic reconnections between the loops thus driving  high-speed outflows along the loops. Velocities of disturbances propagating along the sunspot loops appear to be temperature dependent indicating the slow magneto-acoustic nature of the propagating waves \citep[e.g.,][]{2009A&A...505..791S,2012SoPh..279..427K}. In the corona the oscillations are limited to small regions that often coincide with the endpoints of sunspot coronal loops \citep{2004SoPh..221..237B} and show no signs of the resonances \citep{2002SoPh..207..259B} predicted by the chromospheric resonator theories \citep{1983SoPh...82..369Z, 2011ApJ...728...84B}.

Sunspot oscillations also manifest themselves via umbral flashes \citep[UFs,][]{raey}, which are emissions in the core of chromospheric lines caused by upward propagating waves steepening into shocks \citep{2010ApJ...722..888B}. The temperature in UFs may exceed the surrounding values by up to 1000K and fluctuations of umbral fields are possible on sub-arcsecond scales, although the field inclination and the overall magnetic configurations does not seem to change \citep{2003A&A...403..277R}. \cite{spikes} recently reported on observations of very small scale (height $<$~1Mm, width $<$ 0.1~Mm) jet-like structures in the sunspot umbra that thought to be driven by sunspot oscillations. \cite{2013ApJ...776...56R} suggested may be a sunspot version of short dynamic fibrils  observed in plage areas. Observed density and field oscillations further support the idea of possible slow mode wave propagation in the atmosphere above sunspot’s umbra.

In this study we analyse data from the space based Interface Region Imaging Spectrograph (IRIS) and the New Solar Telescope (NST) operating the Big Bear Solar Observatory (BBSO) to better understand the fine structure of chromospheric umbral oscillations, the structure and evolution of umbral flashes and shocks. 

\section{Data Description}

We analysed NST \citep{goode_nst_2010} and IRIS \citep{iris} data acquired during joint IRIS-BBSO-DST observations on 2 September 2013. This was a medium sit-and-stare IRIS data set with the slit crossing the center of the umbra of the main sunspot in NOAA AR 11836. The IRIS observing run began at 16:39 UT and lasted for about 100 min. The pixel size of the data is 0$''$.167 and the time cadence was 3~s and 12~s for spectral and slit-jaw data, correspondingly. In this study we used calibrated level 2 data available online. The same IRIS data set was used for studying shocks in the upper chromosphere and the TR above the sunspot by \cite{Tian_2014}. We utilized spectra and slit-jaw images obtained using Mg II 2796.35\AA\ and Si IV 1393.76\AA\ spectral lines, which form in the chromosphere (formation temperature 10$^4$~K) and middle of the transition region (TR, 10$^{4.9}$~K), respectively.  This data set also includes series of images from Atmospheric Imaging Assembly on board of Solar Dynamic Observatory (SDO/AIA) \citep{aia}. 

The NST data were acquired with the aid of the 308 sub-aperture adaptive optics (AO-308) and span the same time interval as IRIS observations. Time series of broadband (10\AA) images of the photosphere were taken with a TiO filter (7057\AA, time cadence 15~s, pixel scale 0$''$.0375). This absorption line (the head of the TiO $\gamma$-system) is only formed at low temperatures below 4000~K, and is well suitable for observing sunspot umbra, where the temperatures are significantly lower \citep{2003A&A...412..513B}.

We also utilized available series of narrow-band H$\alpha$ (6563\AA) images taken at $\pm$ 0.8\AA, $\pm$ 0.4\AA, and 0.0\AA\ from the line center acquired with NST's Visible Imaging Spectrometer (VIS, pixel size of 0$''$.029), which combines a 5\AA\ interference filter with a Fabry-P\'{e}rot etalon to produce a resulting bandpass of 0.07\AA\ over a 70$''$ $\times$ 70$''$ field of view. The difference in the acquisition time at two sequential line positions (e.g, +0.4\AA\ and -0.4\AA) was about 2~s. The five point line scan was recorded every 23~s. At each line position we acquired a burst of 25 images with the exposure times ranging from 12~ms at $\pm$0.8\AA\ to 25~ms at the line center. The bursts were then speckle reconstructed \citep{kisip_code}. In \cite{spikes} we have estimated that the errors introduced by local misalignment and residual seeing into measurements at each line position ranges from 2\% to 9\% with the average of 5\%.

We found that VIS H$\alpha$+0.8\AA\ and IRIS Mg II 2796.35\AA\ images are best suitable for co-aligning the NST and IRIS data sets. We estimate the maximum error of co-alignment to be about 0$''$.5 (3 IRIS pixels). The co-alignment was performed manually using prominent features in the umbra and several brightening events outside the sunspot. IRIS images and spectra were co-aligned using the fiducial line on the slit.

\begin{figure}[!ht]
\centering
\begin{tabular}{c}
\epsfxsize=5.5truein  \epsffile{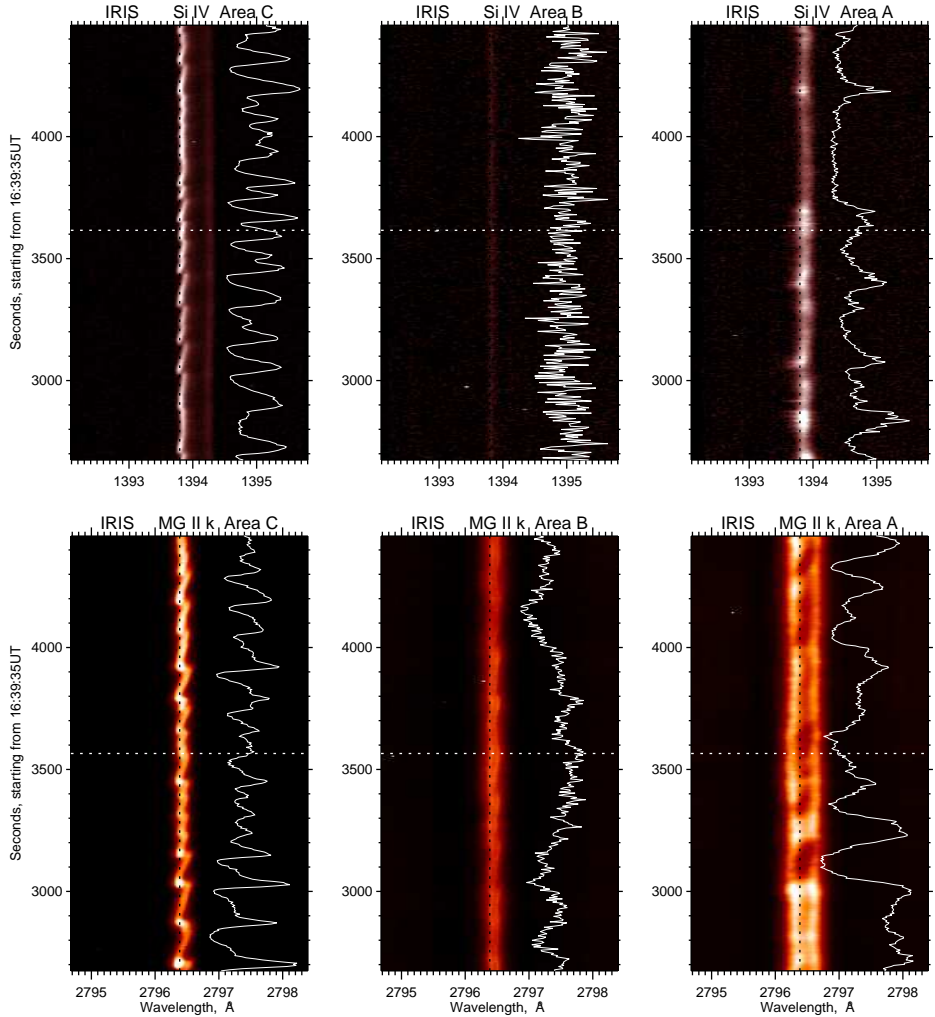} \
\end{tabular}
\caption{\sl Time evolution of IRIS Si IV 1394\AA\ (top) and Mg II k 2796\AA\ (bottom) spectral lines at three locations (area A, B, and C) associated with a sunspot. The white oscillating curves on the right side of the spectra are intensity time profiles as measured along the black dotted line. The horizontal dotted line marks the acquisition time of the images shown in Fig. \ref{nst_iris}.}
\label{iris_sp}
\end{figure}

\section{Results}

\subsection{Sunspot oscillations and shocks} 
\begin{figure}[!ht]
\centering
\begin{tabular}{c}
\epsfxsize=6.5truein  \epsffile{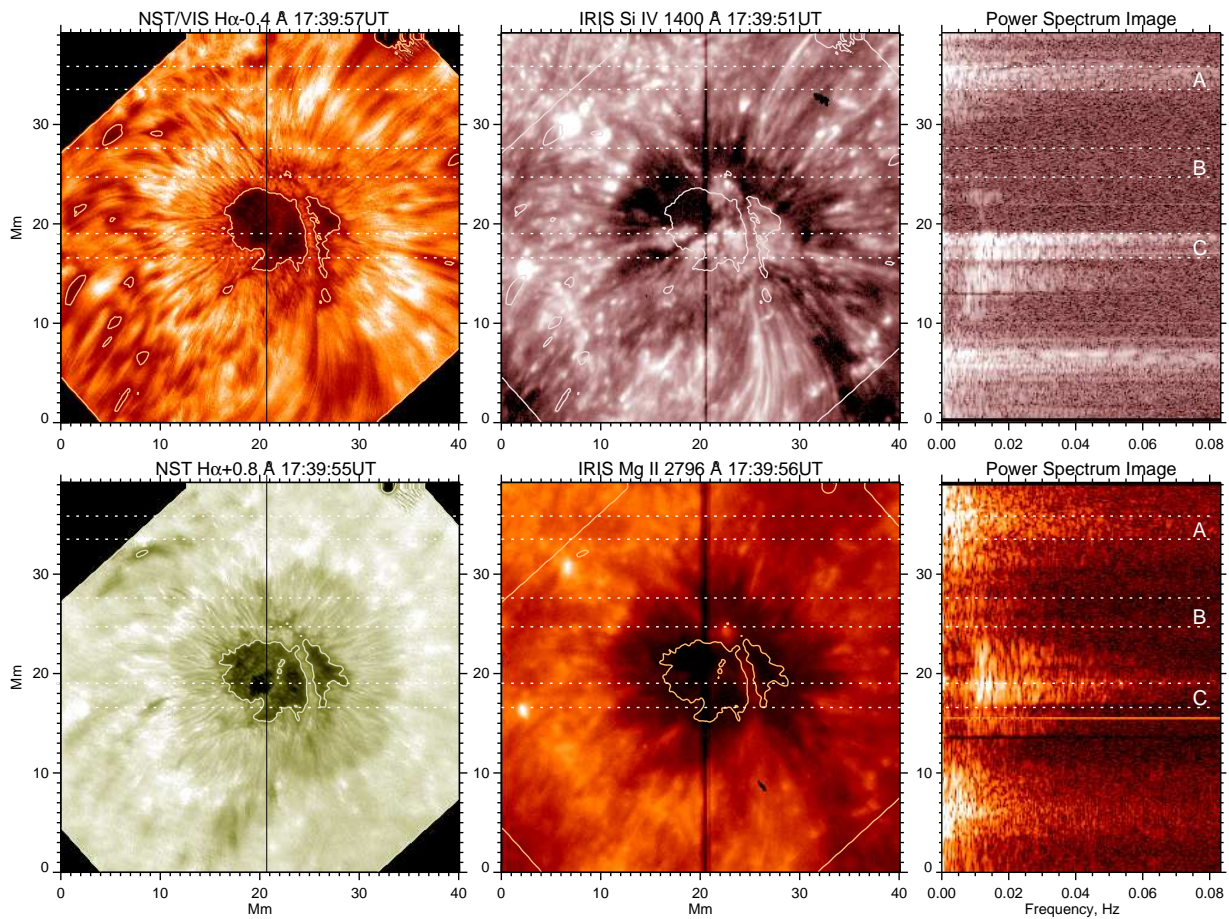} \
\end{tabular}
\caption{\sl NST \ha\ (left), IRIS slit-jaw images (center) and IRIS power spectra images (right). The vertical line across chromospheric images marks the location of the IRIS slit. The three sets of horizontal dotted lines outline areas A, B, and C selected for analysis. The power spectra image is a spatial stack of many power spectra calculated at each pixel position along the IRIS slit using intensity time profiles shown in Fig. \ref{iris_sp} (left). The corresponding spectral plots, averaged over the outlines areas are shown in Fig. \ref{mean_sp}. The contours outline \ha\ umbra and are shown here to ease the comparison.}
\label{nst_iris}
\end{figure}

In Figure \ref{iris_sp} we show time evolution of IRIS Si IV (top) and Mg II k (bottom) spectra for three different positions along the slit (see Figure \ref{nst_iris}). The spectral evolution at area C is mainly associated with the light bridge (LB) in the middle of the umbra, while areas B and A represent the penumbra and a plage area outside the sunspot, respectively. As evident from the figure, the atmosphere above the sunspot is filled with periodic shocks (left panels) of different duration and strength, which are manifested here as the sawtooth pattern in the plot of spectral evolution \citep[also see][]{Tian_2014}. The oscillating curves plotted on the right of the spectra are intensity time profiles measured $\sim$0.05\AA\ off each line center (black vertical dashed lines) that corresponds to the average shift of the spectral line due to a passing shock. The Si IV line above the penumbra (area B) appears to be very weak and does not show presence of shocks. The corresponding Mg II k spectrum (lower middle panel) in the same area does indicate the presence of weak shocks, possibly associated with propagating penumbral waves. The corresponding intensity curve also suggests that there might be long period intensity oscillations (about 15 min). Finally, both lines show no signature of shocks in the plage area A, where the Mg II k line is seen restored back to its double peak form (lower right). The Si IV line shows several features e.g., at t=2800-2900~s, near t=3150~s, and 3400~s, which can be described as rapid blue shifted excursions \citep[see][]{2009ApJ...705..272R}, and therefore can possibly be interpreted as signatures of rapid upflows of plasma associated with, e.g., type II spicules \citep{2007PASJ...59S.655D}.

\begin{figure}[!ht]
\centering
\begin{tabular}{c}
\epsfxsize=5.7truein  \epsffile{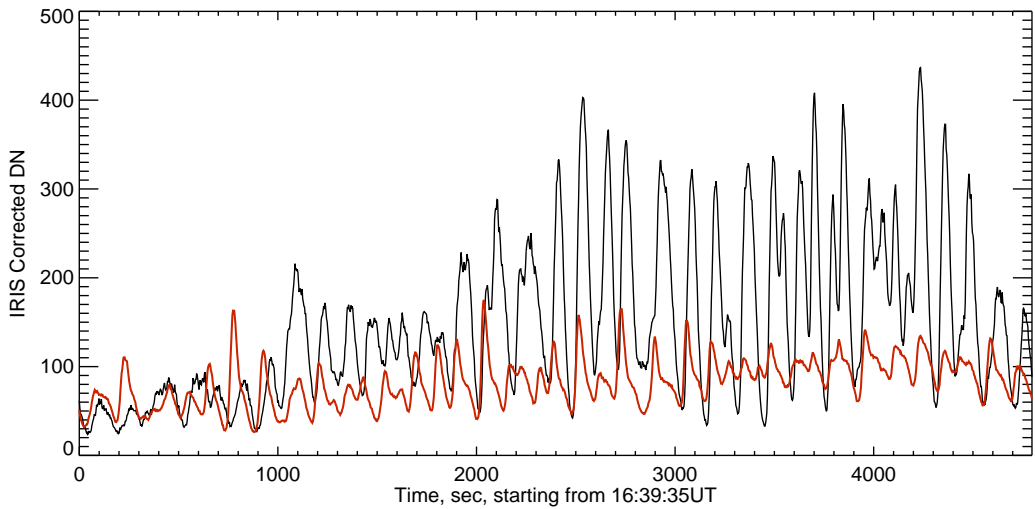} \
\end{tabular}
\caption{\sl Mg II k (red) and Si IV (black) blue wing intensity time profiles measured 0.05\AA\ off the line center along the black dotted line for area C (center of the umbra). Time counting starts at the beginning of the IRIS observing run.}
\label{cross_corr}
\end{figure}

Figure \ref{cross_corr} plots the same profiles as in Figure \ref{iris_sp} but only for the entire observed interval. The corresponding cross-correlation function peaks at t=-15~s indicating that the occurrence of Si IV shocks (black curve) is delayed relative to those of the Mg II k line (red curve), thus indicating that shocks as well as the wave flux propagate upward in the sunspot atmosphere. We also note that this delay time is comparable to that derived from single Gaussian Doppler shift at one slit location by \cite[][see the correction in Madsen et al. 2014]{Tian_2014} and derived from the same IRIS data and for the same sunspot \citep[see also][ for other cross-correlation results]{2013SoPh..284..379K}. The comparison of individual features in the two time profiles indicates that the above derived lag is rather an average value. For example, between t=2400~s and t=2800~s the lag reaches up to 40~s, while it is nearly zero at t=4200 -- 4500~s. At the same time there are intervals where no reasonable correlation between the intensity features can be found (e.g., at t=1200-1800~s). The zig-zag pattern seen in the two left panels of Figure \ref{iris_sp} also supports this conclusion. Thus, Mg II k spectra (chromosphere) between t=3000 -- 3100~s show an ``abnormal$''$ shock pattern, i.e, a sequence of six short duration ($\approx$ 70~s), low amplitude shocks, while the Si IV data (TR) during the same period display the regular amplitude, long duration shocks. The pattern is reversed between t=3800~s and t=4200~s, when the Si IV line shows almost no shock activity in the TR, while the chromosphere was beating with its regular rhythm. These oscillation patterns suggest that a complex interaction may exist between the upward travelling waves and those reflected from the transition region and/or corona back to the photosphere \citep{1982SoPh...75...35H, 2006RSPTA.364..313B, 2011ApJ...728...84B, 2014ApJ...793...43C}. At times, the wave interaction may lead to suppression or enhancement of shocks, depending on the phase of the interacting waves. It is also curious that the Mg II k intensity profile for the penumbra (lower middle panel) seem to exhibit long term (approx. 20~min) intensity variations. We probably only see one cycle of what may be cyclic variations, however, we would like to point out that similar long period oscillations have previously been found in outer penumbra of sunspots \citep{2008ASPC..383..279B,2013A&A...554A.146K}. It is also worth noting that the abnormal Si IV shock pattern in the umbra at t=3800 -- 4200~s (top left panel) seem to coincide in time with the minimum in the presumed long period penumbral wave (bottom center panel), while the Mg II k abnormal umbral shock activity (bottom left) seem to occur when the long period wave have crested. It does, however, remain to be seen whether these long period variations have a well pronounced periodicity. As it follows from Figure \ref{cross_corr} the intensity of the shocks is not particularly stable even over longer periods of time. Thus, for about first 30 min of observations the shocks appear weak, while their strength increased gradually. Between t=2000~s and t=4500~s the shocks, especially in the TR were regular and strong, while their intensity seemed to rapidly decreased after that time interval. It is not readily obvious the nature of such long period variations.

NST H$\alpha$ (left column) and IRIS (center) images of the sunspot are shown in Figure \ref{nst_iris}. Although the H$\alpha$-0.4\AA\ features (top left panel) seen outside the sunspot carry some resemblance to those of the Si IV 1394\AA\ (top center), the umbra appears very different when observed in these two spectral ranges. While the H$\alpha$-0.4\AA\ umbra is split in three parts by two major LBs, the Si IV 1396\AA\ images, show an umbra filled with various bright structures that not even remotely resemble the chromospheric structures. On the other hand, the Mg II k 2796\AA\ slit-jaw images (bottom center) of the umbra do show the LBs. 

The right column in Figure \ref{nst_iris} are images of power spectra of intensity oscillations (each image row represents one power spectrum) calculated as follows. For each pixel along the IRIS slit we generated an intensity profile in the same way as described above and in Figure \ref{iris_sp}. A power spectrum was calculated for each time profile and the spatial stack of the spectra is shown in the right column. There is large scale pattern visible in the power spectra variations across the sunspot. The oscillations (shocks) are most intense in the umbra (y=13-24~Mm), very weak in the penumbra ( y=10-14~Mm and y=23-29~Mm) and they intensify again in the plage outside the sunspot (y$<$10~Mm and y$>$32~Mm). This behavior is similar in both analyzed IRIS spectral lines, although there are some differences. 

\begin{figure}[!ht]
\centering
\begin{tabular}{c}
\epsfxsize=6.5truein  \epsffile{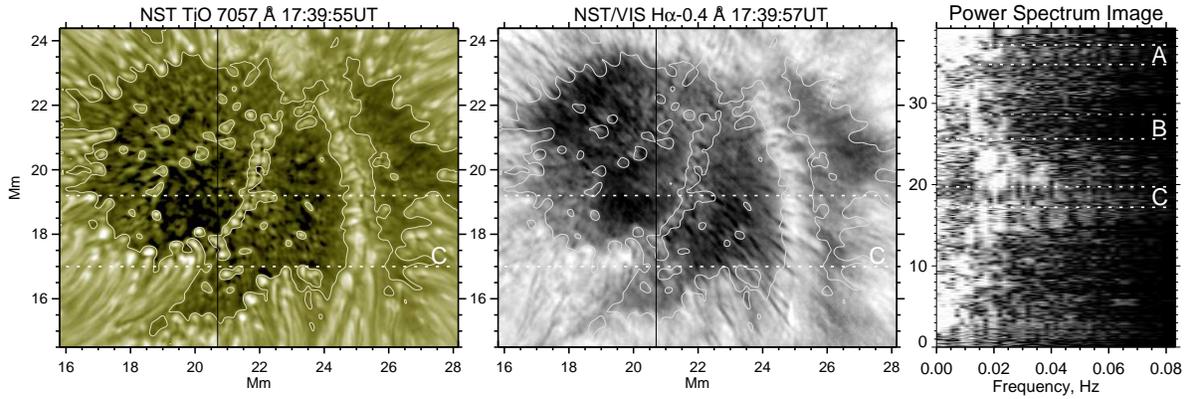} \
\end{tabular}
\caption{\sl NST TiO (photosphere, left) and \ha\ (low chromosphere, middle) high resolution images on the sunspot's umbra. The vertical line across chromospheric images marks the location of the IRIS slit. The two horizontal dotted lines outline the area C. The \ha\ power spectra image (right) was obtained in the same way as the IRIS power spectra image. Note the y-coordinate for the images and the power image are different. The three sets of horizontal lines outline areas A, B, and C. The contours outline TiO bright features at an arbitrary level and are shown here to ease the comparison.}
\label{nst_sp}
\end{figure}

The power spectrum plots indicate that the intensity of shocks inside the umbra varies along the slit as well as with height. Thus, area C (y=16.5-18~Mm) was defined as a part of the umbra with the most intense (steeper power spectrum) and it also is associated with an LB. It is interesting to note the sudden cutoff of Si IV oscillations at y=19~Mm, although the Mg II k line shows that the oscillations gradually decrease as we move away from the center of the umbra. Although in both IRIS slit-jaw images the umbra at y $>$ 19~Mm appear relatively dark, the waves that presumably travel upward from the photosphere are not detected in the TR. The entire second quadrant of the umbra appears dark at the TR although the photospheric and lower chromosphere images shown in Figure \ref{nst_sp} do not show anything unusual about that area. The left and middle panel of this figure are a zoomed-in view of the umbra. The penumbra in the  photospheric (TiO 7305\AA) image (left) appears to be more or less uniformly dotted with umbral dots (UDs) of various size and intensity \citep[see][]{2012ApJ...745..163K}. The photospheric umbra at the location that corresponds to the darkest TR umbra (x$<$20.5~Mm; y$>$19~Mm) appear very similar to a part of the photospheric umbra (x=22-24~Mm; y=17-19~Mm) that is very bright in the TR line. At the same time, the the darkest part of the photospheric umbra (x=19-21~Mm; y=18-20~Mm) appears bright in the TR. 

The middle panel in Figure \ref{nst_sp} is a high resolution image of the chromosphere above the umbra, where the LBs are still visible and some of the UDs can be identified as well. At the same time it is covered with umbral spikes \citep[short dark streaks,][]{spikes}, which are thought to be sunspot’s version of classical spicules and may be driven by sunspot oscillations. Finally the right panel in this figure is a image of H$\alpha$-0.4\AA\ power spectra obtained in the exact manner as the IRIS power spectra discussed above. Although the frequency resolution of the \ha\ power spectra (22~s cadence time series) is inferior in comparison to those of IRIS data (3~s cadence) they also show variations across the umbra similar to those we saw in the Mg II k line.

We will now briefly summarize these figures and conclude that there is large scale pattern in variations of the shock intensity across a sunspot as well as with height above the sunspot. Shocks observed at the blue wing of spectral lines, appear to be more intense above LBs. At the same time, there are locations inside the umbra where TR shocks cannot be detected and we speculate that the magnetic field configuration at those locations may be responsible for shock production and wave propagation above a sunspot.

\begin{figure}[!ht]
\centering
\begin{tabular}{ccc}
\epsfxsize=2.1truein  \epsffile{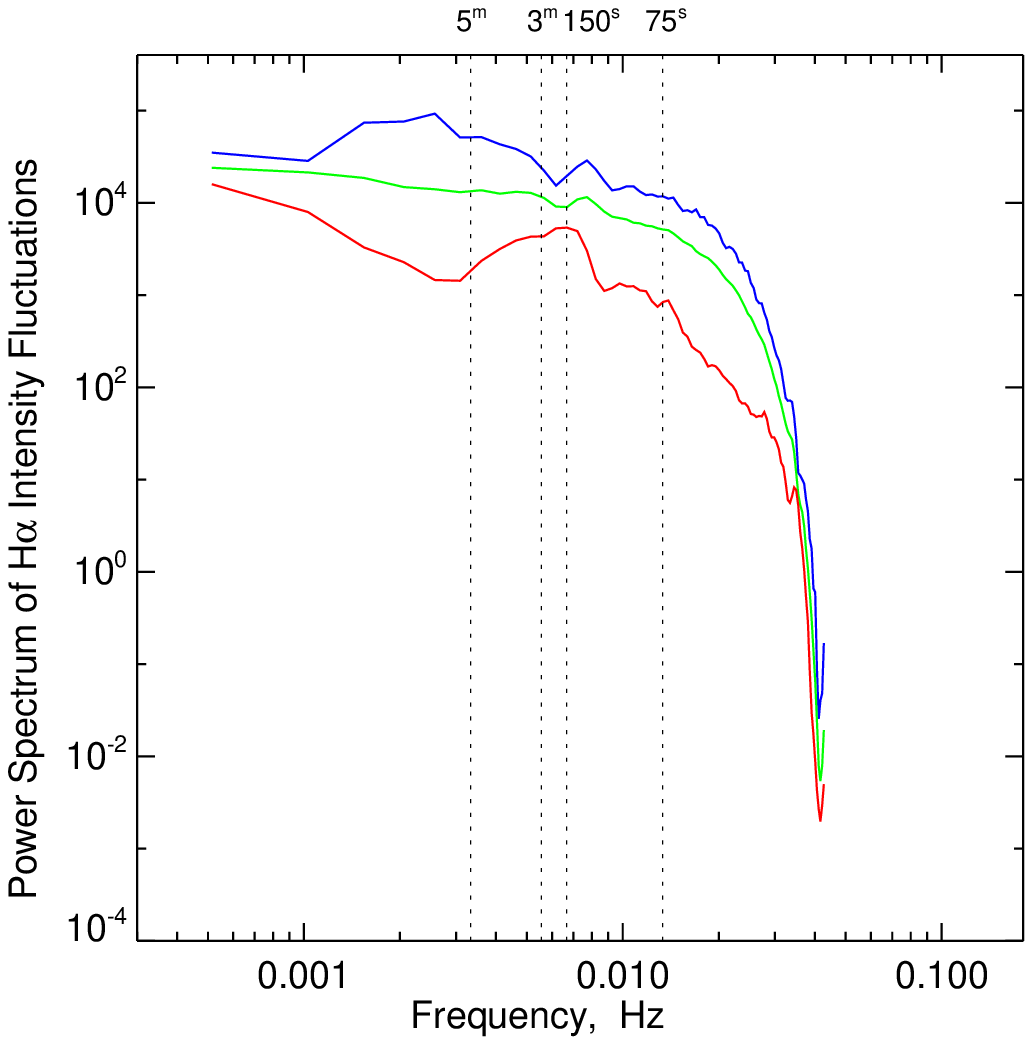} &
\epsfxsize=2.1truein  \epsffile{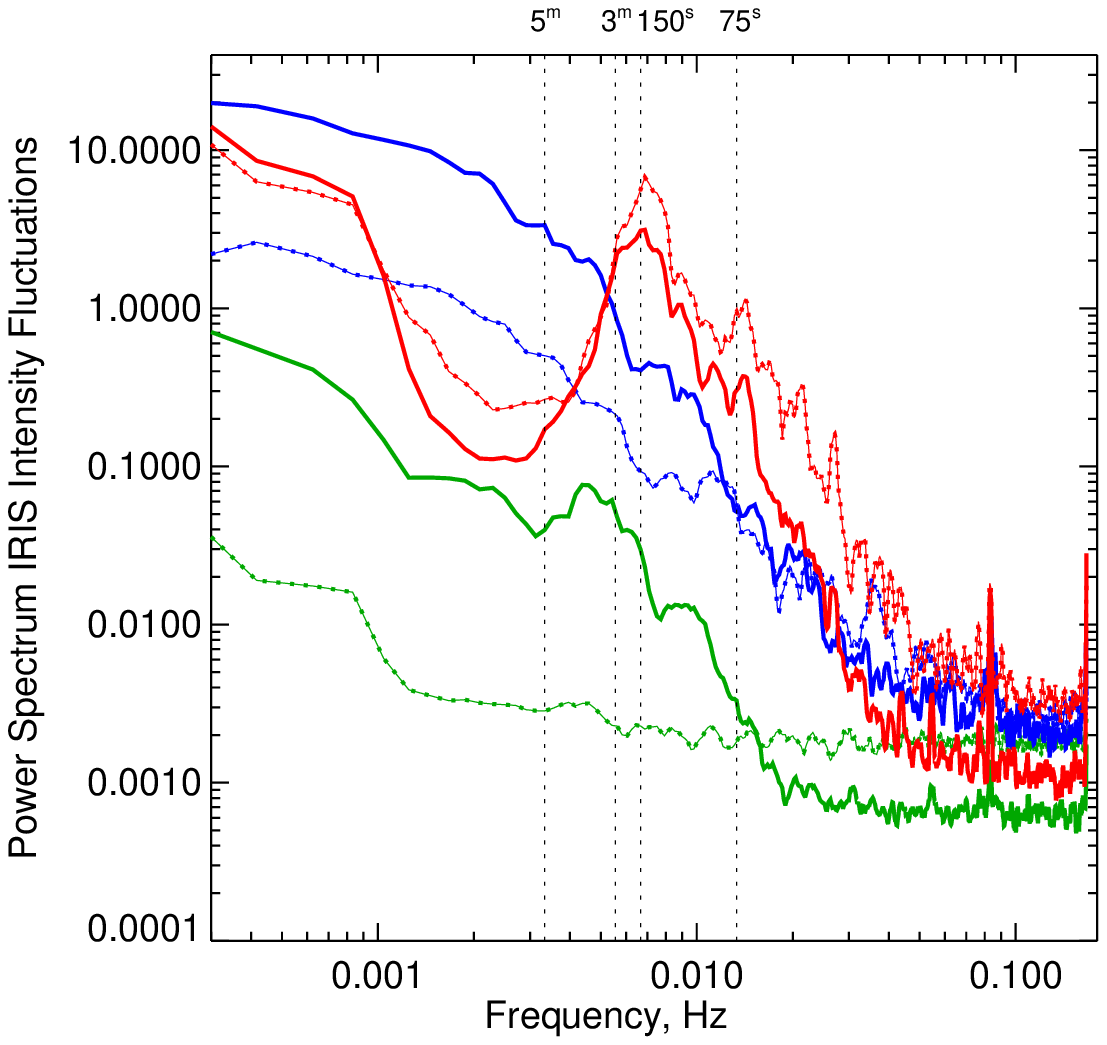} &
\epsfxsize=2.1truein  \epsffile{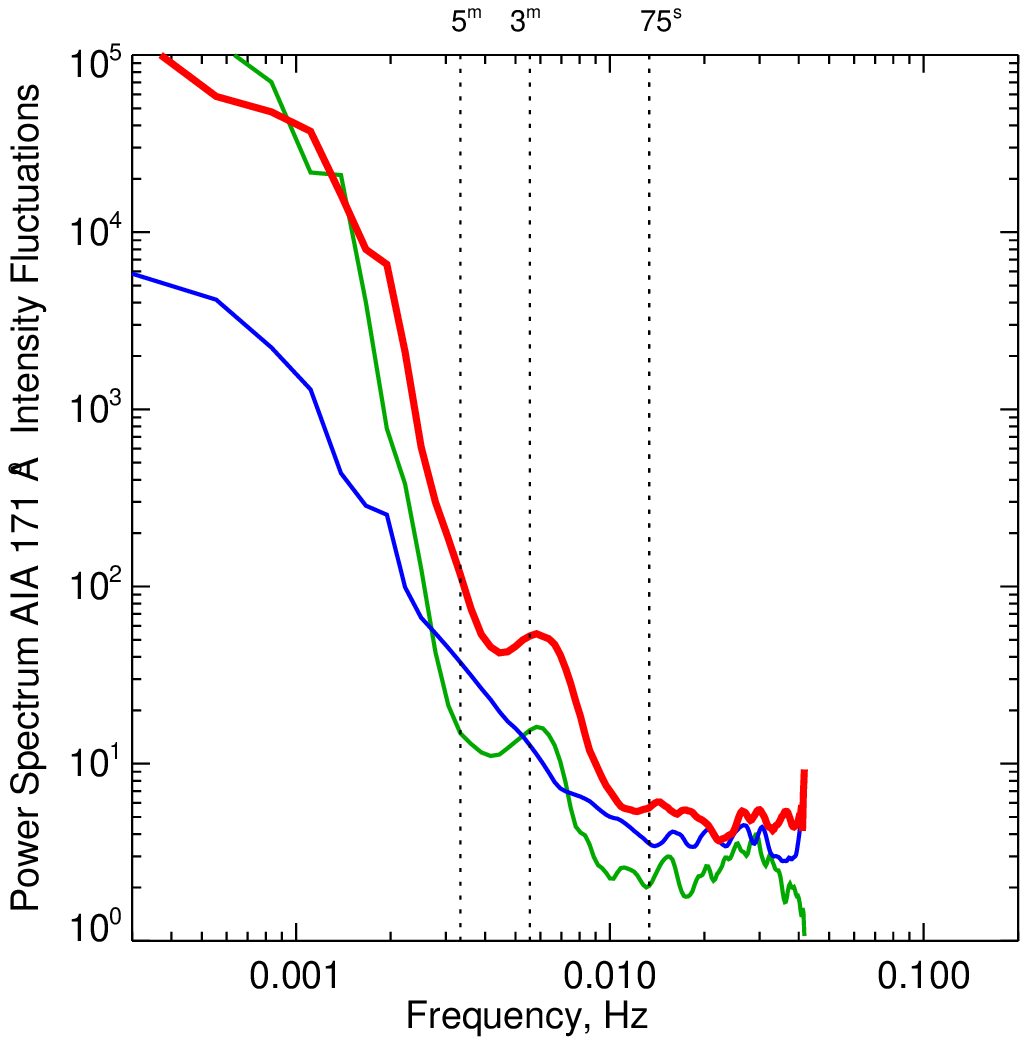} \
\end{tabular}
\caption{\sl Averaged power spectra of \ha\ (left), IRIS (middle), and AIA (right) intensity  oscillations for area A (blue, plage), B (green, penumbra), and C (red, umbra). In the middle panel thick (thin) lines show results for Mg II k (Si IV) spectra.}
\label{mean_sp}
\end{figure}

\begin{figure}[!ht]
\centering
\begin{tabular}{c}
\epsfxsize=4.5truein  \epsffile{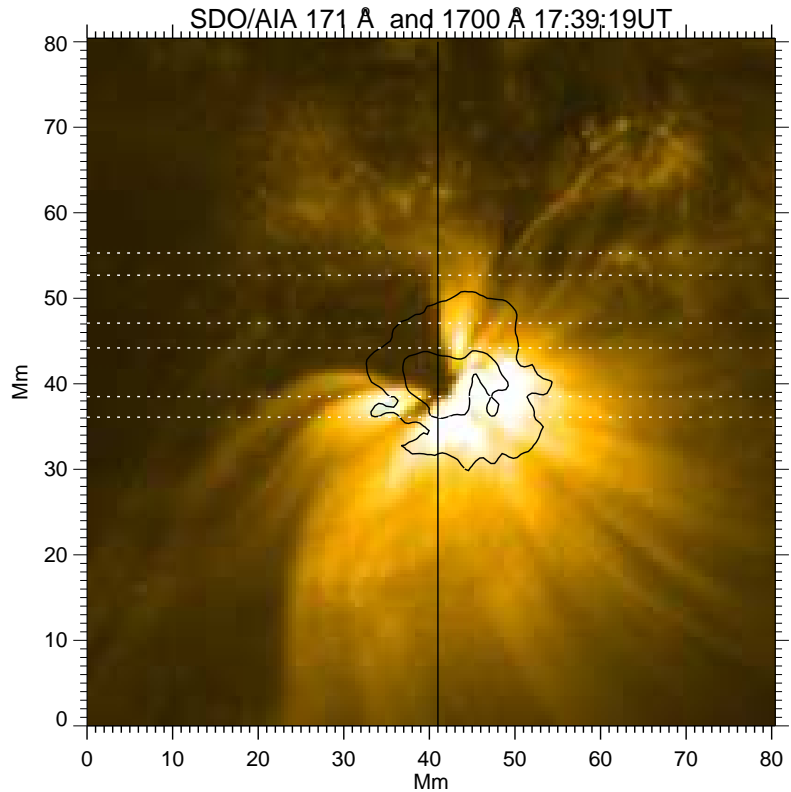} \
\end{tabular}
\caption{\sl SDO AIA 171\AA\ image of the sunspot showing bright legs of coronal loops rooted in the umbra (small contour). The longer contour outlines the sunspot's penumbra. The three sets of horizontal dotted lines outline the areas A, B, and C.}
\label{aia}
\end{figure}

Figure \ref{mean_sp} shows power spectra of intensity oscillations as observed at the \ha\ (left), IRIS (center), and 171\AA\ (right, see also Figure \ref{aia}) lines averaged over the three areas A (blue, plage), B (green, penumbra), and C (red, umbra). The umbral \ha\ (left), Si IV (center, thin), Mg II (center, thick) and the 171\AA\ spectra  all show presence of significant power (i.e., a broad peak with a maximum at around 3~min), while they also display a lack of power of 5~min oscillations.

The penumbral oscillations (running penumbral waves) in the upper chromosphere (Mg II) are less intense as compared to the umbral counterparts and they clearly show that longer period photospheric oscillations (power peak between 5~min and 3~min, solid green) penetrate in the upper atmosphere as the sunspot fields become more inclined \citep[e.g.,][]{2006ApJ...647L..73H}. At the same time, the penumbral waves do not show strong presence in the TR Si IV power spectra (dotted green). In the lower chromosphere (\ha-0.4\AA) the situation reverses in the sense that the penumbral oscillations contain more power than the umbral ones and the spectrum is uniform with few small deviation from a smooth curve. We should note that the penumbral \ha\ spectrum (green) is the only one that does not contain a well defined inertial interval (see for example the red \ha\ and IRIS spectra), which indicates that the detected oscillations are a superposition of several independent oscillation sources (waves), rather than being an energy cascade driven by one main source (e.g., photospheric 5~min).

Both penumbral and umbral 171\AA\ spectra look rather similar with the exception of the total power they represent. The ``penumbral$''$ coronal spectrum (green) shows a 3~min peak, which is drastically different from the Mg II spectra and highlights the fact that due to expansion of umbral fields, the coronal intensity oscillations above the penumbra are driven and related to the sunspot umbra, rather than to the penumbral fields (Figure \ref{aia}). Finally, the plage spectra (blue) measured at all four atmospheric levels are similar as well, having the longest inertial interval, and they do not display any significant and peculiar peaks along the frequency range. 

Figure \ref{aia} shows the corona above the sunspot in the 171\AA\ line as seen using the SDO/AIA instrument. Considering that these images predominantly map legs of coronal loops rooted in sunspots, we may conclude that the brightness structures seen at the TR (Si IV) and the coronal altitudes are very similar. Therefore, the source of heating of the footpoints of coronal loops may be hidden in the TR or even deeper.

\subsection{Structure and Dynamics of Umbral Flashes}

The wave phenomena in the chromosphere of sunspots manifest themselves also as umbral flashes (UFs). We will use NST \ha\ data to take a closer look at the dynamics and the structure of UFs.

\begin{figure}[!ht]
\centering
\begin{tabular}{c}
\epsfxsize=4.5truein  \epsffile{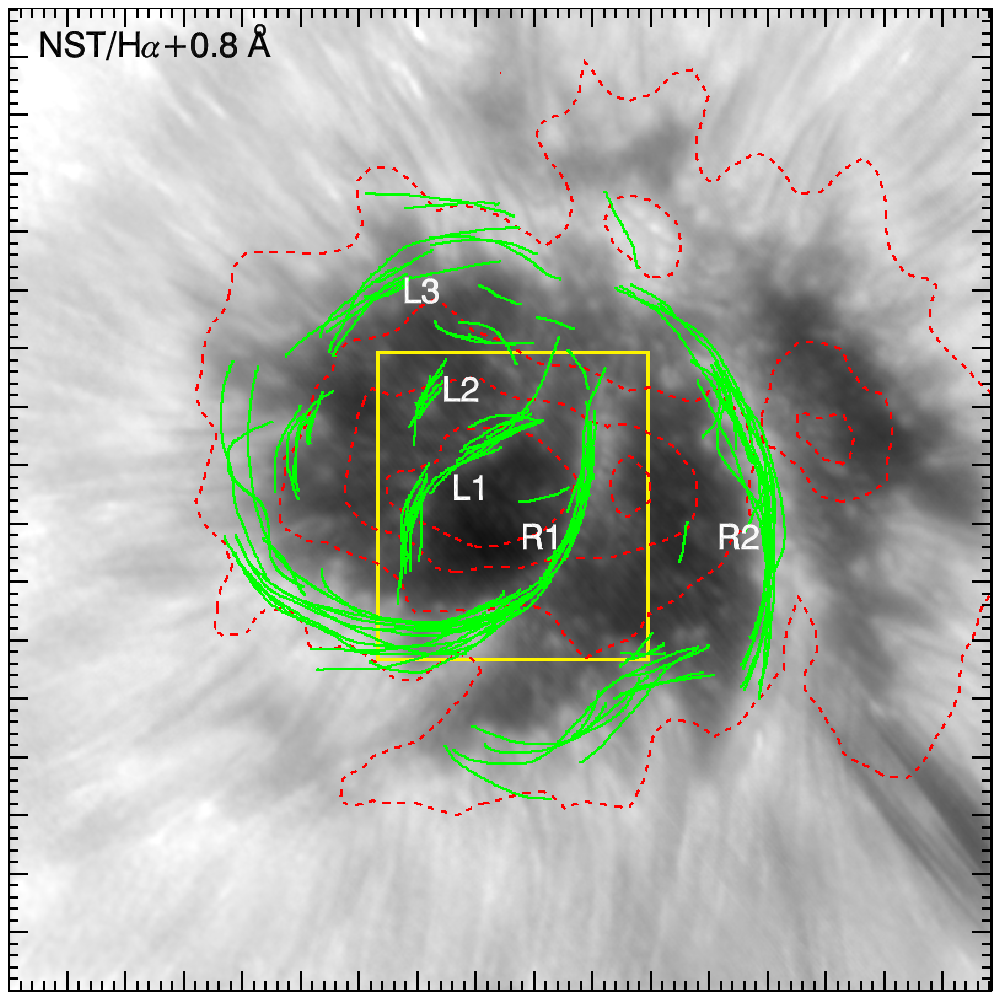} \
\end{tabular}
\caption{\sl Locations of bright lanes of umbral flashes (UFs, greed line segments) plotted over an \ha\ off band image. The red dotted contours indicate HMI line of sight magnetic field strength at -1000, -1400, -1600, and -1800~G. L1, L2, L3, R1, and R2 mark various UFs lanes discussed in the text.}
\label{ufs}
\end{figure}

We traced and plotted all the locations where UFs have appeared (Figure \ref{ufs}, green lines). We find that in this particular case all UFs structures were extended and formed bright lanes of various length, as opposed to being diffuse patches randomly distributed over the umbra. It also turned out that the UFs appearance in the umbra is limited to several well defined locations: i) the umbra-penumbra boundary (e.g., L3), ii) the LBs (L1, R1, and R2) and iii) clusters of bright umbral dots (e.g., L2). In other words, these are locations with enhanced magneto-convection features and possibly larger (as compared to dark umbra) gradients of the magnetic fields, since the umbral dots (UDs) and LBs are thought to be magneto-convection cells that develop in a field free environment. At the edge of these convection cells, umbral field weakens and expands rapidly while the cells themselves rise above the dark umbra. It is interesting to note that the bright UF lanes tend to appear on that side of these raised convective structures that face the center of the umbra. For example, lanes R1 and R2 always appear on the left side of the associated LBs. UFs inside the dark and uniform (no large UDs) parcels of umbra (such as located between L1 and R1) are either weak and diffused, or they do not register at all. 

\begin{figure}[!ht]
\centering
\begin{tabular}{c}
\epsfxsize=6.5truein  \epsffile{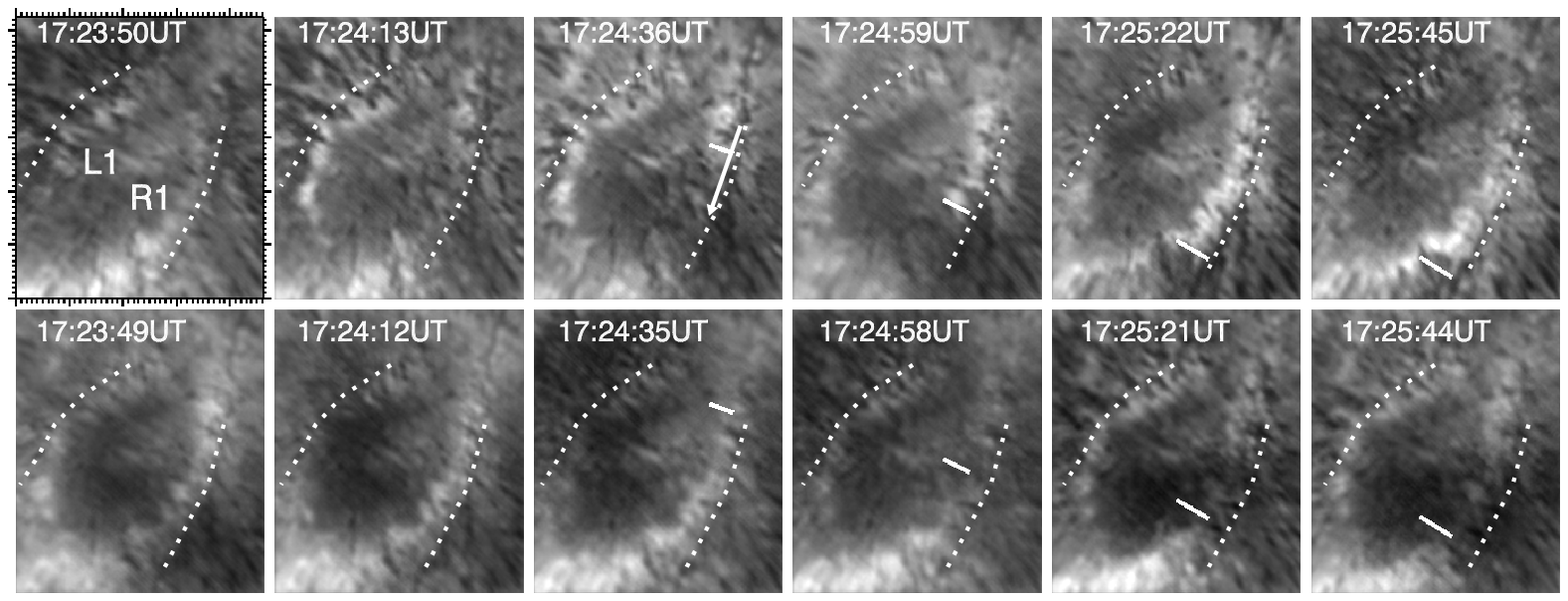} \
\end{tabular}
\caption{\sl Dynamics of UFs as seen in NST \ha+0.4\AA\ (top) and \ha-0.4\AA\ (bottom) images. L1 and R1 mark locations of two bright UF lanes (the same as in Fig. \ref{ufs}). The short line segments mark the edge of an expanding UF lane, while the arrow shows the direction of expansion. The large tick marks in the leftmost top panel indicate 1~Mm intervals.}
\label{ha_ufs}
\end{figure}

Figure \ref{ha_ufs} sheds light on details of spatial distribution and time evolution of UFs inside the yellow box in Figure \ref{ufs}. The top (bottom) row shows a sequence of \ha+0.4\AA~ (\ha-0.4\AA) images that span a nearly 2~min time interval. The beginning of the interval was chosen to be at the moment when the \ha\ line was minimally shifted due to the passing shock. The next blue wing frame (17:24:12~UT) shows that a diffuse darkening has appeared between L1 and R1, closer to L1, while the corresponding red wing frame shows a narrow bright UF lane at that location. In the next frame (17:24:35~UT), the darkening has spread toward the upper left corner of the frame possibly indicating direction of propagation of the subsurface waves driving the shock. The corresponding red wing frame clearly displays umbral spikes surrounded by UFs, which seem to be prominent only when associated with the spikes. The 17:25:21~UT frame presents another cycle of shock development, however, this time the dark cloud spreads from the center toward the lane R1 (i.e., toward the lower right corner of the image). The short line segment in the last four frames mark the edge of an UF lane extending along the umbral light bridge with a rate of about 40km/s. Again, it is worth noting that, at least in this sunspot, bright UFs appeared only in the form of well defined lanes at the edge of a light bridge and around clusters of UDs, while the rest of the umbra seems only weakly disturbed (brightened) by the passing shocks.

\begin{figure}[!ht]
\centering
\begin{tabular}{c}
\epsfxsize=6.5truein  \epsffile{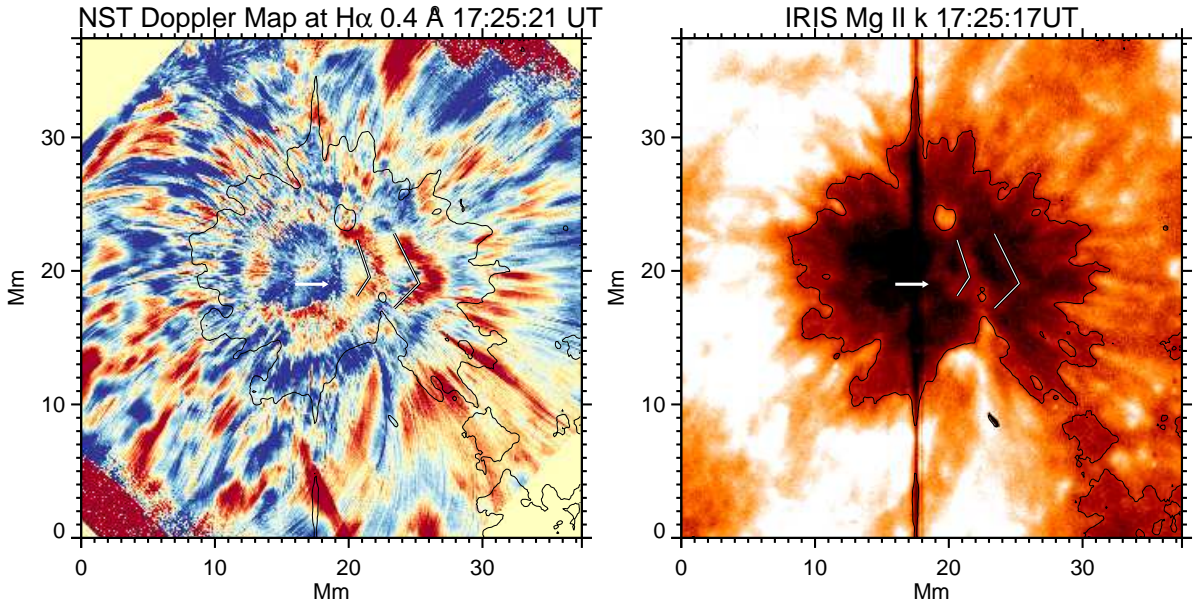} \
\end{tabular}
\caption{\sl NST \ha\ 0.4\AA\ Doppler map (left) and IRIS Mg II k slit-jaw image. The black contour outlines the Mg II k sunspot and is drawn at an arbitrary level.The white arrow points to a blue shifted feature in the \ha\ Doppler map that corresponds to the R1 UF lane. The two broken line segments mark two low contrast Mg II k UF fronts that are co-spatial with the two red-shifted features in the \ha\ Doppler map.}
\label{nst_iris_ufs}
\end{figure}

Finally, in Figure \ref{nst_iris_ufs} we compare an \ha\ Doppler map (left) generated by subtracting an \ha+0.4\AA~ image from the corresponding blue wing image to the nearly co-temporal IRIS Mg II k slit-jaw image (right). The purpose of this comparison is to make a connection between the UFs seen in \ha\ and IRIS data. The arrows in both panels indicate the location of the UF lane discussed in Figure \ref{ha_ufs}, while the two broken lines mark the position of diffuse Mg II k UFs. Although these Mg II k UFs are not high contrast features they were clearly identified from a movie of the slit-jaw images. It turns out that the Mg II k flashes are co-spatial with locations of red-shifted phase of \ha\ shock waves, i.e., times when the shocked plasma was falling down and the \ha\ line was red-shifted. Note, that the IRIS UFs are somewhat delayed relative to the corresponding \ha\ features, which can be explained by the time difference between the two images. In the next IRIS image taken at 17:25:29~UT the UFs are already further away from the sunspot center and better coincide with the \ha\ features, however, their contrast is too low to be reproduced in a figure. We note that the two red-shifted \ha\ arches and the Mg II k UFs marked by the broken lines are co-spatial with the R2 arch in Figure \ref{ufs} and the umbra-penumbra boundary, respectively. Moreover, similarly to \ha\ UFs, the Mg II k flashes, when observed away from the light bridges, appear much weaker and harder to track, thus further confirming our earlier suggestion about possible importance of the LBs in formation and/or enhancement of shocks.

\section{Conclusions and Discussions}

IRIS and NST spectral data further confirm the idea of upward propagating magneto acoustic waves, which has been reported earlier by a number of observers. The average time lag between the arrival times of the shocks at the Mg II k and Si IV line formation levels, determined from a cross correlation analysis of 100~min long time profiles of blue wing intensities, was found to be about 15~s. We find, however, that this lag is highly dependent on the selection of the correlated segments and peak to peak delay may range from 40~s to zero. Analyzing a 100~min long time profile we find that the intensity of chromospheric shocks above the sunspot's umbra may vary on large time scales of order of 20~min. However, the available data set was obviously too short to make more confident conclusions.

These long period waves may have internal or external (relative to a sunspot) origin. One may speculate that the modulation may arise from a hysteresis effect, where the current physical conditions of the shocked plasma are defined not only by the currently propagating shock but also by the previous history of plasma heated by passing shocks. As shocks constantly push the chromosphere up, not all the plasma may timely fall back to the initial state by the time the next one arrives, which may lead to a temporarily slightly over-inflated chromosphere. When the critical mass that can no longer be supported by shock waves is accumulated, it falls back suppressing the upwardly moving shocks \citep[e.g.,][]{1982SoPh...75...35H,Tian_2014,2014ApJ...789..108C}.  An alternative external cause for the irregular beat may be the sub-photospheric interaction of the upward propagating fast magneto-acoustic waves and those reflected at the photosphere back into the convective zone as discussed by \cite{0004-637X-719-1-357}. These effects may have a global influence on the sunspot causing 18-24~min oscillations in Evershed flows \citep{2008ASPC..383..279B}, 18~min periodicities at the chromospheric layers that were concentrated at the outer penumbra \citep{2013A&A...554A.146K},  chromospheric flows above sunspot \cite[e.g.,][]{flocculent}, as well as upward wave propagation. 

NST's \ha\ data showed that bright UFs prefer to be located near LBs and clusters of UDs. This is a new and significant finding, indicating that although magneto-acoustic waves are present everywhere inside the umbra, the effectiveness of UF production may depend on the non-uniformity in the umbra's magneto-convection properties and magnetic fields. This inference agrees with \cite{2007A&A...463.1153T} conclusion that there are umbral areas with slightly different physical and/or magnetic field conditions. This preference for UFs to be linked to LBs and UD clusters may be explained as follows. The UDs and LBs are known to have weaker magnetic fields \citep{1993A&A...278..584W, 2004ApJ...614..448S, 2004ApJ...604..906R, 2008ApJ...672..684R,2009ApJ...702.1048W, 2012ApJ...757...49W} as compared to the dark umbra dotted only with small, dim and barely discernible UDs. Therefore they are thought to be surrounded by casp-like fields with a horizontal field gradient \citep{2009ApJ...704L..29L}. \cite{1982SoPh...75...35H} found that the shock formation depends on the vertical gradient of Alfven speed, $v_A$, in the medium. In particular,  if $v_A$, decreases with height then shocks form more effectively. Such conditions are met either in rapidly expanding vertical flux tubes and/or near edges of flux tubes, where fields may become mostly horizontal. In sunspot umbra, such conditions exist near LBs and UD clusters, which are considered field-free structures. In this case adjacent umbral fields expand into the atmosphere above these features thus creating a narrow lane of negative vertical gradient of $v_A$ with favourable conditions for formation of chromospheric shocks. 

The observations show that UFs tend to appear on the sunspot-center side of LBs. Such dynamics suggests that they may be linked to a presumed horizontal trans-umbral wave that causes UFs to sequentially appear at progressively distant light bridges. According to \cite{0004-637X-719-1-357} simulations of magneto-acoustic waves in sunspots using a 3D nonlinear MHD code, we speculate that we might be dealing with a compact sub-photospheric driver of oscillations as opposed to photospheric oscillations being coherent on the scales of the entire umbra. \cite{0004-637X-719-1-357} showed that when a fast (acoustic) wave moves horizontally across the umbral field lines, up and away from the subphotospheric source located near the axis of a sunspot, it is partially converted into a slow mode wave when it reaches the layer where the Alfven and sound speeds are equal. The slow mode waves then steepen into shocks thus creating a pattern of shocks (UFs) propagating away from the sunspot center. This behavior inferred from simulations is very similar to the dynamics of UFs observed here.

Most of the sunspot umbra, observed in the TR Si IV spectral line, appears bright and filled with oscillations and shocks. However, there are regions that appear very dark at the chromospheric and TR level and do not exhibiting detectable oscillations. While other authors have pointed out the connection between the footpoints of coronal loops and the bright TR umbra \cite[e.g., ][]{Tian_2014}, we emphasize a possible connection between the bright magneto-convection features (i.e., LBs, clusters of UDs) and bright UV umbra and coronal loop footpoints. To the best of our knowledge, the role of LBs in coronal heating has not been widely discussed in the science literature. We find that the bright Si IV umbra is situated above the underlying chromospheric LBs and other bright umbral structures as well as with the footpoints of the overlying bright coronal loops. The intensity of shocks is not uniform over the umbra and the most intense shock structures were detected in the parts of the umbra associated with one of the LBs. We suggest that the non-homogeneities in the umbral magnetic field may play an important role in heating coronal loops rooted in the sunspots by facilitating generation of shocks. The non-uniformities may also increase effectiveness of fast to slow mode conversion, however this question has to be answered by future MHD simulations. 

Finally, as we have already noted in this study, the second quadrant of the umbra appears dark in Si IV and AIA 171\AA\ (as well as other) spectral lines. Careful inspection of the relevant AIA and BSBO \ha\ images showed that the dark part of the umbra was associated with a system of what appears to be dim horizontal threads possibly representing a filaments channel. \cite{2014KPCB...30...40K} used spectropolarimetric data acquired at the Fe I 15643 - 15658\AA\ to conclude that in the presence of moderate and strong magnetic fields (300 - 1800~G) the photospheric waves tend to run upward, while in the areas with low closed loops oscillations exhibited a lack of power and standing waves dominate. 

While these observations offer reasonable explanations for existence and spatial distribution of hot and bright UV loops and dim umbra, we still need to understand the mechanisms of wave energy transfer and conversion, heating of umbral loops and the role of light bridges in this process. 

This work was conducted as part of the effort of NASA's Living with a Star Focused Science Team ``Jets''. We thank BBSO observing and engineering staff for support and observations. This research was supported by NASA LWS NNX11AO73G and NSF AGS-1146896 grants. VYu acknowledges support from Korea Astronomy and Space Science Institute during his stay there, where a part of the work was performed. Authors thank Prof. Nakaryakov for insightful comments. We also thank anonymous referee for valuable criticism. 


\begin{thebibliography}{47}
\expandafter\ifx\csname natexlab\endcsname\relax\def\natexlab#1{#1}\fi

\bibitem[{{Balasubramaniam} {et~al.}(2008){Balasubramaniam}, {Pevtsov}, \&
  {Olmschenk}}]{2008ASPC..383..279B}
{Balasubramaniam}, K.~S., {Pevtsov}, A.~A., \& {Olmschenk}, S. 2008, in
  Astronomical Society of the Pacific Conference Series, Vol. 383, Subsurface
  and Atmospheric Influences on Solar Activity, ed. R.~{Howe}, R.~W. {Komm},
  K.~S. {Balasubramaniam}, \& G.~J.~D. {Petrie}, 279

\bibitem[{{Bard} \& {Carlsson}(2010)}]{2010ApJ...722..888B}
{Bard}, S., \& {Carlsson}, M. 2010, \apj, 722, 888

\bibitem[{Beckers \& Tallant(1969)}]{raey}
Beckers, J., \& Tallant, P. 1969, Solar Physics, 7, 351

\bibitem[{{Berdyugina} {et~al.}(2003){Berdyugina}, {Solanki}, \&
  {Frutiger}}]{2003A&A...412..513B}
{Berdyugina}, S.~V., {Solanki}, S.~K., \& {Frutiger}, C. 2003, \aap, 412, 513

\bibitem[{{Bogdan}(2000)}]{2000SoPh..192..373B}
{Bogdan}, T.~J. 2000, \solphys, 192, 373

\bibitem[{{Bogdan} \& {Judge}(2006)}]{2006RSPTA.364..313B}
{Bogdan}, T.~J., \& {Judge}, P.~G. 2006, Royal Society of London Philosophical
  Transactions Series A, 364, 313

\bibitem[{{Botha} {et~al.}(2011){Botha}, {Arber}, {Nakariakov}, \&
  {Zhugzhda}}]{2011ApJ...728...84B}
{Botha}, G.~J.~J., {Arber}, T.~D., {Nakariakov}, V.~M., \& {Zhugzhda}, Y.~D.
  2011, \apj, 728, 84

\bibitem[{{Brynildsen} {et~al.}(2004){Brynildsen}, {Maltby}, {Foley},
  {Fredvik}, \& {Kjeldseth-Moe}}]{2004SoPh..221..237B}
{Brynildsen}, N., {Maltby}, P., {Foley}, C.~R., {Fredvik}, T., \&
  {Kjeldseth-Moe}, O. 2004, \solphys, 221, 237

\bibitem[{{Brynildsen} {et~al.}(2002){Brynildsen}, {Maltby}, {Fredvik}, \&
  {Kjeldseth-Moe}}]{2002SoPh..207..259B}
{Brynildsen}, N., {Maltby}, P., {Fredvik}, T., \& {Kjeldseth-Moe}, O. 2002,
  \solphys, 207, 259

\bibitem[{{Chae} {et~al.}(2014){Chae}, {Yang}, {Park}, {Ajor Maurya}, {Cho}, \&
  {Yurchysyn}}]{2014ApJ...789..108C}
{Chae}, J., {Yang}, H., {Park}, H., {Ajor Maurya}, R., {Cho}, K.-S., \&
  {Yurchysyn}, V. 2014, \apj, 789, 108

\bibitem[{{Chmielewski} {et~al.}(2014){Chmielewski}, {Murawski}, {Musielak}, \&
  {Srivastava}}]{2014ApJ...793...43C}
{Chmielewski}, P., {Murawski}, K., {Musielak}, Z.~E., \& {Srivastava}, A.~K.
  2014, \apj, 793, 43

\bibitem[{{De Moortel}(2005)}]{2005RSPTA.363.2743D}
{De Moortel}, I. 2005, Royal Society of London Philosophical Transactions
  Series A, 363, 2743

\bibitem[{{De Moortel} {et~al.}(2002){De Moortel}, {Ireland}, {Hood}, \&
  {Walsh}}]{2002A&A...387L..13D}
{De Moortel}, I., {Ireland}, J., {Hood}, A.~W., \& {Walsh}, R.~W. 2002, \aap,
  387, L13

\bibitem[{{de Pontieu} {et~al.}(2007){de Pontieu}, {McIntosh}, {Hansteen},
  {Carlsson}, {Schrijver}, {Tarbell}, {Title}, {Shine}, {Suematsu}, {Tsuneta},
  {Katsukawa}, {Ichimoto}, {Shimizu}, \& {Nagata}}]{2007PASJ...59S.655D}
{de Pontieu}, B., {et~al.} 2007, \pasj, 59, 655

\bibitem[{{De Pontieu} {et~al.}(2014){De Pontieu}, {Title}, {Lemen}, {Kushner},
  {Akin}, {Allard}, {Berger}, {Boerner}, {Cheung}, {Chou}, {Drake}, {Duncan},
  {Freeland}, {Heyman}, {Hoffman}, {Hurlburt}, {Lindgren}, {Mathur}, {Rehse},
  {Sabolish}, {Seguin}, {Schrijver}, {Tarbell}, {W{\"u}lser}, {Wolfson},
  {Yanari}, {Mudge}, {Nguyen-Phuc}, {Timmons}, {van Bezooijen}, {Weingrod},
  {Brookner}, {Butcher}, {Dougherty}, {Eder}, {Knagenhjelm}, {Larsen},
  {Mansir}, {Phan}, {Boyle}, {Cheimets}, {DeLuca}, {Golub}, {Gates}, {Hertz},
  {McKillop}, {Park}, {Perry}, {Podgorski}, {Reeves}, {Saar}, {Testa}, {Tian},
  {Weber}, {Dunn}, {Eccles}, {Jaeggli}, {Kankelborg}, {Mashburn}, {Pust},
  {Springer}, {Carvalho}, {Kleint}, {Marmie}, {Mazmanian}, {Pereira}, {Sawyer},
  {Strong}, {Worden}, {Carlsson}, {Hansteen}, {Leenaarts}, {Wiesmann},
  {Aloise}, {Chu}, {Bush}, {Scherrer}, {Brekke}, {Martinez-Sykora}, {Lites},
  {McIntosh}, {Uitenbroek}, {Okamoto}, {Gummin}, {Auker}, {Jerram}, {Pool}, \&
  {Waltham}}]{iris}
{De Pontieu}, B., {et~al.} 2014, \solphys, 289, 2733

\bibitem[{Felipe {et~al.}(2010)Felipe, Khomenko, \&
  Collados}]{0004-637X-719-1-357}
Felipe, T., Khomenko, E., \& Collados, M. 2010, The Astrophysical Journal, 719,
  357

\bibitem[{{Goode} {et~al.}(2010){Goode}, {Coulter}, {Gorceix}, {Yurchyshyn}, \&
  {Cao}}]{goode_nst_2010}
{Goode}, P.~R., {Coulter}, R., {Gorceix}, N., {Yurchyshyn}, V., \& {Cao}, W.
  2010, Astronomische Nachrichten, 88, 789

\bibitem[{{Hansteen} {et~al.}(2006){Hansteen}, {De Pontieu}, {Rouppe van der
  Voort}, {van Noort}, \& {Carlsson}}]{2006ApJ...647L..73H}
{Hansteen}, V.~H., {De Pontieu}, B., {Rouppe van der Voort}, L., {van Noort},
  M., \& {Carlsson}, M. 2006, \apjl, 647, L73

\bibitem[{{Hollweg} {et~al.}(1982){Hollweg}, {Jackson}, \&
  {Galloway}}]{1982SoPh...75...35H}
{Hollweg}, J.~V., {Jackson}, S., \& {Galloway}, D. 1982, \solphys, 75, 35

\bibitem[{{Kiddie} {et~al.}(2012){Kiddie}, {De Moortel}, {Del Zanna},
  {McIntosh}, \& {Whittaker}}]{2012SoPh..279..427K}
{Kiddie}, G., {De Moortel}, I., {Del Zanna}, G., {McIntosh}, S.~W., \&
  {Whittaker}, I. 2012, \solphys, 279, 427

\bibitem[{{Kilcik} {et~al.}(2012){Kilcik}, {Yurchyshyn}, {Rempel}, {Abramenko},
  {Kitai}, {Goode}, {Cao}, \& {Watanabe}}]{2012ApJ...745..163K}
{Kilcik}, A., {Yurchyshyn}, V.~B., {Rempel}, M., {Abramenko}, V., {Kitai}, R.,
  {Goode}, P.~R., {Cao}, W., \& {Watanabe}, H. 2012, \apj, 745, 163

\bibitem[{{Kobanov} {et~al.}(2013{\natexlab{a}}){Kobanov}, {Kolobov}, {Kustov},
  {Chupin}, \& {Chelpanov}}]{2013SoPh..284..379K}
{Kobanov}, N., {Kolobov}, D., {Kustov}, A., {Chupin}, S., \& {Chelpanov}, A.
  2013{\natexlab{a}}, \solphys, 284, 379

\bibitem[{{Kobanov} {et~al.}(2013{\natexlab{b}}){Kobanov}, {Chelpanov}, \&
  {Kolobov}}]{2013A&A...554A.146K}
{Kobanov}, N.~I., {Chelpanov}, A.~A., \& {Kolobov}, D.~Y. 2013{\natexlab{b}},
  \aap, 554, A146

\bibitem[{{Kobanov} \& {Makarchik}(2004)}]{2004A&A...424..671K}
{Kobanov}, N.~I., \& {Makarchik}, D.~V. 2004, \aap, 424, 671

\bibitem[{{Kostik} {et~al.}(2014){Kostik}, {Shchukina}, {Kobanov}, \&
  {Pulyaev}}]{2014KPCB...30...40K}
{Kostik}, R.~I., {Shchukina}, N.~G., {Kobanov}, N.~I., \& {Pulyaev}, V.~A.
  2014, Kinematics and Physics of Celestial Bodies, 30, 40

\bibitem[{{Lemen} {et~al.}(2012){Lemen}, {Title}, {Akin}, {Boerner}, {Chou},
  {Drake}, {Duncan}, {Edwards}, {Friedlaender}, {Heyman}, {Hurlburt}, {Katz},
  {Kushner}, {Levay}, {Lindgren}, {Mathur}, {McFeaters}, {Mitchell}, {Rehse},
  {Schrijver}, {Springer}, {Stern}, {Tarbell}, {Wuelser}, {Wolfson}, {Yanari},
  {Bookbinder}, {Cheimets}, {Caldwell}, {Deluca}, {Gates}, {Golub}, {Park},
  {Podgorski}, {Bush}, {Scherrer}, {Gummin}, {Smith}, {Auker}, {Jerram},
  {Pool}, {Soufli}, {Windt}, {Beardsley}, {Clapp}, {Lang}, \& {Waltham}}]{aia}
{Lemen}, J.~R., {et~al.} 2012, \solphys, 275, 17

\bibitem[{{Louis} {et~al.}(2009){Louis}, {Bellot Rubio}, {Mathew}, \&
  {Venkatakrishnan}}]{2009ApJ...704L..29L}
{Louis}, R.~E., {Bellot Rubio}, L.~R., {Mathew}, S.~K., \& {Venkatakrishnan},
  P. 2009, \apjl, 704, L29

\bibitem[{{Maurya} {et~al.}(2013){Maurya}, {Chae}, {Park}, {Yang}, {Song}, \&
  {Cho}}]{2013SoPh..288...73M}
{Maurya}, R.~A., {Chae}, J., {Park}, H., {Yang}, H., {Song}, D., \& {Cho}, K.
  2013, \solphys, 288, 73, http://arxiv.org/abs/1302.1234

\bibitem[{{Nakariakov} {et~al.}(2000){Nakariakov}, {Verwichte}, {Berghmans}, \&
  {Robbrecht}}]{2000A&A...362.1151N}
{Nakariakov}, V.~M., {Verwichte}, E., {Berghmans}, D., \& {Robbrecht}, E. 2000,
  \aap, 362, 1151

\bibitem[{{Rimmele}(2008)}]{2008ApJ...672..684R}
{Rimmele}, T. 2008, \apj, 672, 684

\bibitem[{{Rimmele}(2004)}]{2004ApJ...604..906R}
{Rimmele}, T.~R. 2004, \apj, 604, 906

\bibitem[{{Rouppe van der Voort} \& {de la Cruz
  Rodr{\'{\i}}guez}(2013)}]{2013ApJ...776...56R}
{Rouppe van der Voort}, L., \& {de la Cruz Rodr{\'{\i}}guez}, J. 2013, \apj,
  776, 56

\bibitem[{{Rouppe van der Voort} {et~al.}(2009){Rouppe van der Voort},
  {Leenaarts}, {de Pontieu}, {Carlsson}, \& {Vissers}}]{2009ApJ...705..272R}
{Rouppe van der Voort}, L., {Leenaarts}, J., {de Pontieu}, B., {Carlsson}, M.,
  \& {Vissers}, G. 2009, \apj, 705, 272

\bibitem[{{Rouppe van der Voort} {et~al.}(2003){Rouppe van der Voort},
  {Rutten}, {S{\"u}tterlin}, {Sloover}, \& {Krijger}}]{2003A&A...403..277R}
{Rouppe van der Voort}, L.~H.~M., {Rutten}, R.~J., {S{\"u}tterlin}, P.,
  {Sloover}, P.~J., \& {Krijger}, J.~M. 2003, \aap, 403, 277

\bibitem[{{Socas-Navarro} {et~al.}(2004){Socas-Navarro}, {Mart{\'{\i}}nez
  Pillet}, {Sobotka}, \& {V{\'a}zquez}}]{2004ApJ...614..448S}
{Socas-Navarro}, H., {Mart{\'{\i}}nez Pillet}, V., {Sobotka}, M., \&
  {V{\'a}zquez}, M. 2004, \apj, 614, 448

\bibitem[{{Su} {et~al.}(2013){Su}, {Liu}, {Liu}, {Zhang}, {Zhao}, {Xu}, \&
  {Xie}}]{2013ApJ...762...42S}
{Su}, J.~T., {Liu}, Y., {Liu}, S., {Zhang}, Y.~Z., {Zhao}, H., {Xu}, H.~Q., \&
  {Xie}, W.~B. 2013, \apj, 762, 42

\bibitem[{{Sych} {et~al.}(2009){Sych}, {Nakariakov}, {Karlicky}, \&
  {Anfinogentov}}]{2009A&A...505..791S}
{Sych}, R., {Nakariakov}, V.~M., {Karlicky}, M., \& {Anfinogentov}, S. 2009,
  \aap, 505, 791

\bibitem[{{Tian} {et~al.}(2014){Tian}, {DeLuca}, {Reeves}, {McKillop}, {De
  Pontieu}, {Mart{\'{\i}}nez-Sykora}, {Carlsson}, {Hansteen}, {Kleint},
  {Cheung}, {Golub}, {Saar}, {Testa}, {Weber}, {Lemen}, {Title}, {Boerner},
  {Hurlburt}, {Tarbell}, {Wuelser}, {Kankelborg}, {Jaeggli}, \&
  {McIntosh}}]{Tian_2014}
{Tian}, H., {et~al.} 2014, \apj, 786, 137

\bibitem[{{Tsiklauri} \& {Nakariakov}(2001)}]{2001A&A...379.1106T}
{Tsiklauri}, D., \& {Nakariakov}, V.~M. 2001, \aap, 379, 1106

\bibitem[{{Tziotziou} {et~al.}(2007){Tziotziou}, {Tsiropoula}, {Mein}, \&
  {Mein}}]{2007A&A...463.1153T}
{Tziotziou}, K., {Tsiropoula}, G., {Mein}, N., \& {Mein}, P. 2007, \aap, 463,
  1153

\bibitem[{{Vissers} \& {Rouppe van der Voort}(2012)}]{flocculent}
{Vissers}, G., \& {Rouppe van der Voort}, L. 2012, \apj, 750, 22

\bibitem[{{Watanabe} {et~al.}(2012){Watanabe}, {Bellot Rubio}, {de la Cruz
  Rodr{\'{\i}}guez}, \& {Rouppe van der Voort}}]{2012ApJ...757...49W}
{Watanabe}, H., {Bellot Rubio}, L.~R., {de la Cruz Rodr{\'{\i}}guez}, J., \&
  {Rouppe van der Voort}, L. 2012, \apj, 757, 49

\bibitem[{{Watanabe} {et~al.}(2009){Watanabe}, {Kitai}, \&
  {Ichimoto}}]{2009ApJ...702.1048W}
{Watanabe}, H., {Kitai}, R., \& {Ichimoto}, K. 2009, \apj, 702, 1048

\bibitem[{{Wiehr} \& {Degenhardt}(1993)}]{1993A&A...278..584W}
{Wiehr}, E., \& {Degenhardt}, D. 1993, \aap, 278, 584

\bibitem[{{W{\"o}ger} \& {von der L{\"u}he}(2007)}]{kisip_code}
{W{\"o}ger}, F., \& {von der L{\"u}he}, O. 2007, Appl. Opt., 46, 8015

\bibitem[{{Yurchyshyn} {et~al.}(2014){Yurchyshyn}, {Abramenko}, {Kosovichev},
  \& {Goode}}]{spikes}
{Yurchyshyn}, V., {Abramenko}, V., {Kosovichev}, A., \& {Goode}, P. 2014, \apj,
  787, 58

\bibitem[{{Zhugzhda} {et~al.}(1983){Zhugzhda}, {Locans}, \&
  {Staude}}]{1983SoPh...82..369Z}
{Zhugzhda}, I.~D., {Locans}, V., \& {Staude}, J. 1983, \solphys, 82, 369

\end{thebibliography}

\end{document}